\documentclass[11pt,twocolumn,superscriptaddress]{revtex4-1}   
\pdfoutput=1
\usepackage{amssymb}
\usepackage{amsmath}
\usepackage{graphicx}
\usepackage{natbib}
\usepackage{mathrsfs}
\usepackage{ulem}
\usepackage{hyperref}
\usepackage[T1]{fontenc}
\usepackage[letterpaper,textwidth=7in,top=.75in,bottom=.75in]{geometry}
\usepackage{color}
\usepackage[dvipsnames]{xcolor}

\usepackage{soul}

\newcommand{\an}[1]{{\color{purple}{#1}}}

\begin{document}
\title{Estimating the degree of non-Markovianity using variational quantum circuits}
\author{Hossein T. Dinani}
\email[]{htdinani@gmail.com}
\address{
Centro de Investigaci\'{o}n DAiTA Lab, Facultad de Estudios Interdisciplinarios,\\ Universidad Mayor, Santiago, Chile}
\address{
 Escuela Data Science, Facultad de Ciencias, Ingenier\'{i}a  y Tecnolog\'{i}a, Universidad Mayor, Santiago, Chile}
\author{Diego Tancara}
\address{Centro de Óptica e Información Cuántica, Universidad Mayor, Santiago, Chile}
\author{Felipe F. Fanchini}
\address{Faculdade de Ciencias, UNESP - Universidade Estadual Paulista, Bauru, SP, 17033-360, Brazil}
\author{Ariel Norambuena}
\address{Universidad Mayor, Vicerrector\'ia de Investigación, Santiago, Chile}
\author{Ra\'ul Coto}
\address{
Department of Physics, Florida International University, Miami, Florida 33199, USA}
\address{Universidad Bernardo O~Higgins, Santiago de Chile, Chile}

\date{\today}
\begin{abstract}
Several applications of quantum machine learning (QML) rely on a quantum measurement followed by training algorithms using the measurement outcomes. However, recently developed QML models, such as variational quantum circuits (VQCs), can be implemented directly on the state of the quantum system (quantum data). Here, we propose to use a qubit as a probe to estimate the degree of non-Markovianity of the environment. Using VQCs, we find an optimal sequence of qubit-environment interactions that yield accurate estimations of the degree of non-Markovianity for the amplitude damping, phase damping, and the combination of both models. We introduce a problem-based ansatz that optimizes upon the probe qubit and the interaction time with the environment. This work contributes to practical quantum applications of VQCs and delivers a feasible experimental procedure to estimate the degree of non-Markovianity. 
\end{abstract}
\maketitle

\section{Introduction}
The last few years have seen a tremendous advance in machine learning (ML) techniques for analyzing data in a wide variety of fields. Quantum physics has also benefited from ML in various aspects such as control of quantum systems, classification and estimation tasks \cite{Dunjko, Sarma, Marquardt, Bukov, Paris, Papic}. In such cases, ML techniques have been used to analyze classical data, obtained from measuring quantum systems. On the other hand, considerable research has been done to take advantage of the quantum properties to improve machine learning techniques \cite{Schuld2015,Biamonte}. The development of quantum artificial neural networks \cite{Beer} and quantum kernel methods \cite{Schuld_arxiv21} are examples of this. 

Towards quantum machine learning algorithms, learning circuits have proved to be a practical approach \cite{Mitarai}. Considering the currently available noisy intermediate-scale quantum computers \cite{Preskill} with few qubits (50-100 qubits), hybrid quantum-classical algorithms have been designed to develop short-depth quantum circuits with free control parameters. These circuits have been termed as variational quantum circuits (VQCs) \cite{McClean2016, Benedetti, Cerezo, Bharti}. In VQCs, the optimization task is done over quantum (free parameters in the quantum circuit) and classical parameters (used in postprocessing) using classical optimization techniques \cite{McClean2016}. 

One of the main obstacles in quantum technologies is the interaction of the quantum system with its surrounding environment which results in the loss of coherence of the quantum system \cite{Breuerbook}. Simplifications are generally imposed on the physical processes. For instance, the so-called Markovian approximation, in which it is assumed that the evolution of the system does not depend on the history of its dynamics, but only on its current state. Thus, memory aspects are ignored, which often works as a good approximation.

However, it is important to emphasize that non-Markovian signatures frequently appear in the dynamics of quantum systems \cite{BreuerRevModPhys,DeVega}. Furthermore, some physical processes are strongly subjected to non-Markovianity, such as reservoir engineering \cite{Bylicka,Addis}, state teleportation \cite{Laine14}, quantum metrology \cite{Chin}, and even current quantum computers \cite{Morris, White}. Moreover, non-Markovianity can be harnessed as a resource \cite{Berk}.

Accurate determination of the degree of non-Markovianity requires a large number of measurements. Moreover, for the measure of non-Markovianity based on entanglement dynamics, an ancillary qubit, protected from interactions with the environment, needs to be considered. In order to surpass these challenges, ML techniques such as neural networks \cite{Luchnikov}, support vector machines \cite{Fanchini}, random forest regressor \cite{Shrapnel}, tensor network-based machine learning \cite{Guo}, and polynomial regression \cite{Goswami} have been used to determine the degree of non-Markovianity of a quantum process. In addition, it has been shown that quantum circuits can be used to simulate non-Markovian dynamics \cite{Garcia-Perez, Head-Marsden} which is an important advance in the study of realistic situations. 

In the present work, we show that the degree of quantum non-Markovianity can be estimated directly from measurements performed on the qubit. Using VQCs, through supervised learning, we find an optimal sequence of qubit-environment interactions to estimate the degree of non-Markovianity, reaching high precision. We apply this methodology to the paradigmatic amplitude and phase damping channels and the combination of both. For each damping channel, we will use exactly solvable models that provides a theoretical framework and parameter range to characterize non-Markovianity.

The remainder of this paper is organized as follows. In Sec.~\ref{open_dynamics_sec} we provide the theoretical framework for the open quantum dynamics and non-Markovianity. Section~\ref{QVC_sec} contains a brief description of variational quantum circuits. In Sec.~\ref{regg_sec} we focus on the estimation of the degree of non-Markovianity of the quantum processes. We conclude the paper in Sec.~\ref{conc_sec}.

\section{Open quantum system dynamics}\label{open_dynamics_sec}
In what follows, we describe two paradigmatic mechanisms for simulating open quantum systems, namely amplitude and phase damping channels.  

\subsection{Amplitude damping}\label{sec:AD}
For the amplitude damping (AD) channel, we consider a qubit interacting with a bath of harmonic oscillators, given by the Hamiltonian ($\hbar=1$) \cite{Hakkika, Whalen}
\begin{eqnarray}
&&H=\omega_0 \sigma_{+} \sigma_{-}+\sum_{k}{\omega_k a^{\dagger}_k a_k} \nonumber\\
&&\qquad+\sum_k (g^{*}_k \sigma_{+}a_k+g_k \sigma_{-}a^{\dagger}_k).
\end{eqnarray}
Here, $\sigma_{+}=\sigma^{\dagger}_{-}=|1 \rangle\langle 0|$ with $|1\rangle$ ($|0\rangle$) corresponding to the excited (ground) state of the qubit with transition frequency $\omega_0$, $a_k (a^{\dagger}_k)$ is the annihilation (creation) operator of the $k$-th mode of the bath with frequency $\omega_k$, and $g_k$ is the coupling between the qubit and the $k$-th mode. We assume that the bath has a Lorentzian spectral density 
\begin{equation}
    J(\omega)=\frac{1}{2\pi}\frac{\gamma_0 \lambda^2}{(\omega_0-\omega)^2+\lambda^2},
\end{equation}
where $\lambda \approx 1/\tau_r$ with $\tau_r$ being the environment correlation time, $\gamma_0\approx 1/\tau_s$ where $\tau_s$ is the typical time scale of the system.

\begin{figure}[t!]
\centering
    \includegraphics[width=0.45\textwidth]{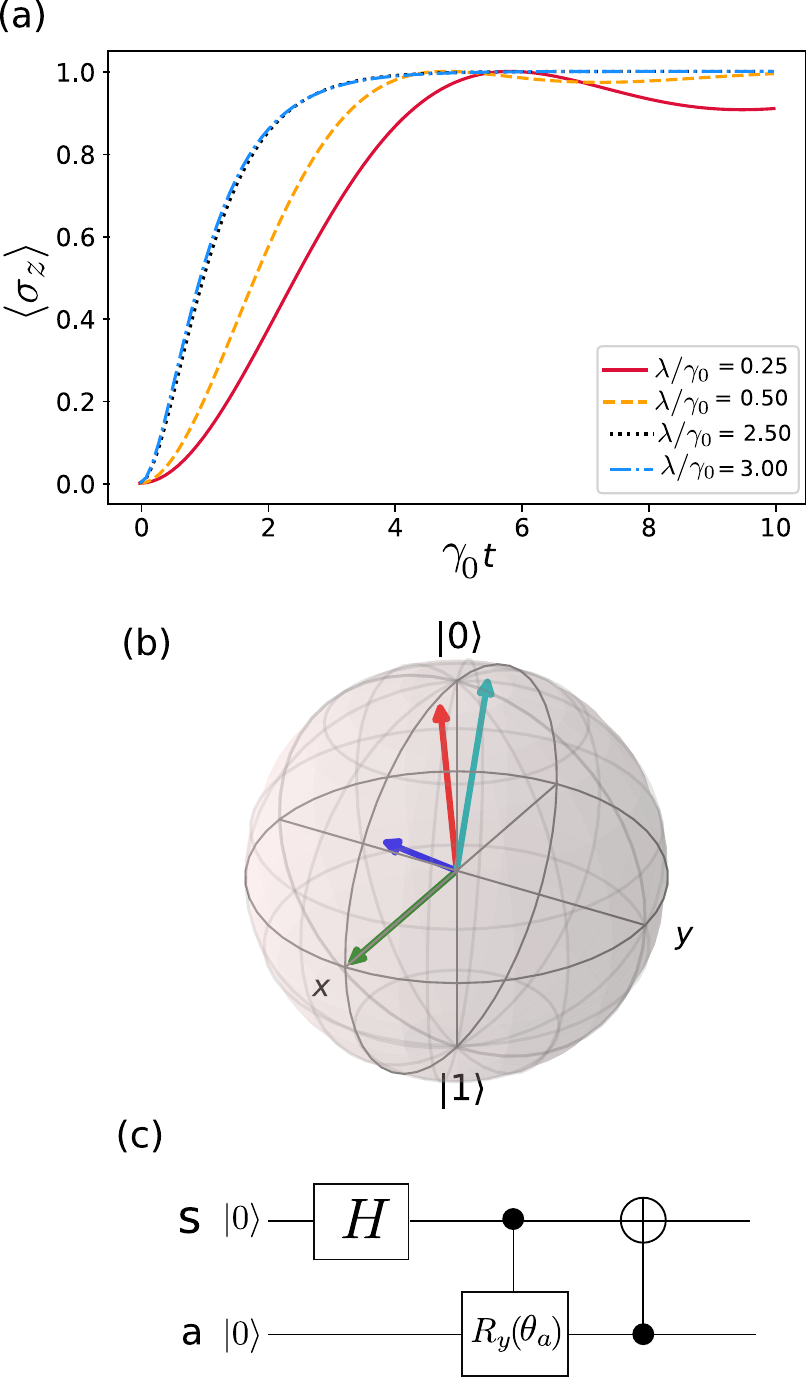}
   \caption{Amplitude damping (AD): (a) The expectation value of the Pauli operator $\sigma_z$ versus $\gamma_0 t$ for a range of values of $\lambda/\gamma_0$. (b) Evolution of the Bloch vector. The qubit is initialized to $\left(|0\rangle+|1\rangle\right)/\sqrt{2}$ and undergoes AD with $\lambda/\gamma_0=0.25$. The Bloch vectors shown are at $\gamma_0 t=0$ (green), $\gamma_0 t=2.5$ (blue), $\gamma_0 t=5$ (red), $\gamma_0 t=10$ (cyan). The Bloch sphere is plotted using QuTiP \cite{qutip}. (c) Quantum circuit for simulating AD.}
   \label{fig:AD3}
\end{figure}

The evolved density matrix for the AD channel can be expressed as follow (super index $a$ refers to \textit{amplitude})
\begin{equation}
    \rho^{(a)}(t)=\sum_{i=0}^{1}{K^{(a)}_i(t)\rho(0)K^{(a)\dagger}_i(t)}, 
\end{equation}
where the Kraus operators are given by \cite{Nielsen, Garcia-Perez}
\begin{eqnarray}\label{eq:kraus_AD}
   && K^{(a)}_0(t)=\begin{pmatrix} 1 & 0 \\ 0 & \sqrt{p_a (t)} \end{pmatrix},\nonumber\\
    &&K^{(a)}_1(t)=\begin{pmatrix} 0 & \sqrt{1-p_a (t)} \\ 0 &0 \end{pmatrix}
\end{eqnarray}
in which \cite{Bellomo}
\begin{equation}\label{eq:ptAD}
    p_a (t)=e^{-\lambda t}\left[\frac{\lambda}{d}\sin\left({d t \over 2} \right)+\cos\left({d t \over 2}\right)\right]^2,
\end{equation}
with $d=\sqrt{2\gamma_0\lambda-\lambda^2}$. The dynamics is known to be non-Markovian in the strong coupling regime $\lambda<2\gamma_0$ $(\tau_s<2\tau_r)$ \cite{Bellomo}. The non-Markovianity can be evidenced in Fig.~\ref{fig:AD3}(a) by the oscillations in the expectation value of the Pauli $z$ operator. Figure~\ref{fig:AD3}(b) shows the evolution of the Bloch vector initialized along the $x$ axis. This representation of the time evolution of the state provides insights for choosing the ansatz for the VQC. For the case of AD, as time evolves the length of the Bloch vector becomes smaller than 1 and after sufficiently long time the vector decays to the ground state $|0\rangle$ recovering its full length. This type of decoherence is also known as longitudinal relaxation which plays a fundamental role in color centers in diamond such as nitrogen-vacancy center coupled to lattice vibrations \cite{Ariel, Matt}, and superconducting qubits in circuit QED \cite{Blais}.

The AD process can be simulated for a general scenario with a quantum circuit via an ancilla qubit \cite{Nielsen,Garcia-Perez}. After tracing out the ancilla qubit\an{,} we obtain the desired mixed state. Figure \ref{fig:AD3}(c) shows the quantum circuit. The Hadamard gate prepares the qubit in the superposition state $\left(|0\rangle+|1\rangle\right)/\sqrt{2}$ while the controlled $y$ rotation and CNOT gates simulate the interaction of the qubit with the environment. In this circuit, the angle $\theta_a$ is given by \cite{Nielsen, Garcia-Perez}
\begin{equation}\label{eq:thetaad}
    \theta_a=2\arccos\left(\sqrt{p_a (t)}\right),
\end{equation}
where $p_a (t)$ is given in Eq.~\eqref{eq:ptAD}

\subsection{Phase damping}\label{sec:PD}
For the phase damping (PD) channel, following Ref.~\cite{Daffer}, we consider a qubit undergoing decoherence induced by a colored noise given by the  stochastic Hamiltonian ($\hbar=1$)
\begin{equation}
H(t)=\Gamma(t) \sigma_z.
\end{equation}
Here, $\Gamma(t)$ is a random variable which obeys the statistics of a random telegraph signal defined as $\Gamma(t)=\alpha (-1)^{n(t)}$, where $\alpha$ is the coupling between the qubit and the external influences, $n(t)$ is a random variable with Poisson distribution with mean $t/(2\tau)$, and $\sigma_z$ is the Pauli $z$ operator. In this case, the dynamics of the qubit is given by the following Kraus operators \cite{Daffer,Nielsen} (super index $p$ refers to \textit{phase})
\begin{eqnarray}\label{eq:kraus_PD}
    &&K^{(p)}_0(t)=\sqrt{\frac{1+\Lambda(t)}{2}}\mathbb{I}, \nonumber\\ 
    &&K^{(p)}_1(t)=\sqrt{\frac{1-\Lambda(t)}{2}}\sigma_z,
\end{eqnarray}
where
\begin{equation}\label{eq:LambdaPD}
\Lambda(t)=e^{-t/(2\tau)}\left[\cos\left(\frac{\mu t}{2\tau}\right)+\frac{1}{\mu}\sin\left(\frac{\mu t}{ 2\tau}\right)\right],
\end{equation}
with $\mu=\sqrt{(4 \alpha \tau)^2-1}$, and $\mathbb{I}$ being the identity matrix. 

\begin{figure}[t!]
\centering
    \includegraphics[width=0.45\textwidth]{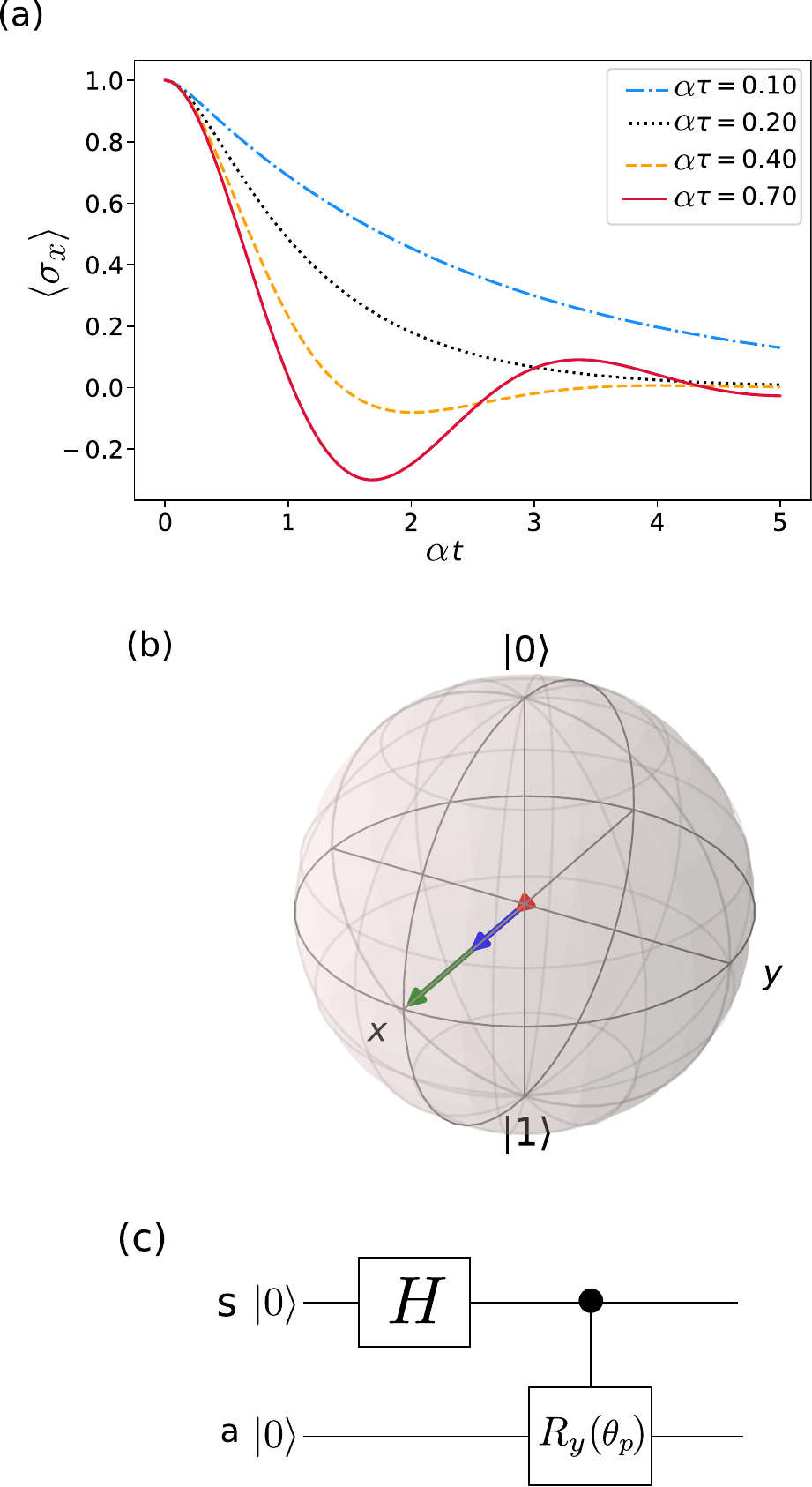}
   \caption{Phase damping (PD): (a) The expectation value of the Pauli operator $\sigma_x$ versus $\alpha t$ for a range of values of $\alpha\tau$. (b) Evolution of the Bloch vector. The qubit is initialized to $\left(|0\rangle+|1\rangle\right)/\sqrt{2}$ and undergoes PD with $\alpha \tau=0.2$. The Bloch vectors shown are at $\alpha t=0$ (green), $\alpha t=1$ (blue), and $\alpha t=5$ (red). (c) Quantum circuit for simulating PD.}
   \label{fig:PD3}
\end{figure}

For $\alpha \tau>1/4$ the dynamics is non-Markovian, while for $\alpha \tau<1/4$ it is Markovian (see Fig.~\ref{fig:PD3}(a)). We note that, in a PD channel, the off diagonal elements of the density matrix decay exponentially as depicted in Fig.~\ref{fig:PD3}(a), where only in the cases $\alpha \tau = \{0.4, 0.7\}$ we observe oscillations. We remind that $\langle\sigma_x\rangle = \langle 0 |\rho |1\rangle +\langle 1 |\rho |0\rangle$ for a two-level system where $\rho$ is the density matrix and $|0\rangle$, $|1\rangle$ are the qubit states. Figure \ref{fig:PD3}(b) shows the evolution of the Bloch vector initialized along the $x$ axis. For such initialization, after sufficiently long time, the vector length decays to zero. In other words, the probability of qubit states is conserved but the phase information between them is lost. Phase damping is also known as transverse relaxation. Examples of systems that undergo this type of decoherence are nitrogen-vacancy center due to its interaction with lattice vibrations \cite{ArielPRA20} and surrounding nuclear spins \cite{deLang}, and superconducting qubits under low-frequency noise \cite{Paladino}.

The PD channel can also be simulated using a quantum circuit, as shown in Fig.~\ref{fig:PD3}(c) \cite{Nielsen}. In this circuit, the Hadamard gate prepares the qubit into the superposition state and the controlled $y$ rotation simulates the interaction with the environment. The angle $\theta_p$ is given by
\begin{equation}\label{eq:thetapd}
    \theta_p=2\arccos\left(\Lambda(t)\right),
\end{equation}
where $\Lambda(t)$ is given in Eq.~\eqref{eq:LambdaPD}.

\subsection{Combined amplitude and phase damping}
There are several system where phase and amplitude damping effects are present at the same time. This generate a more complex scenario from the point of view of non-Markovianity, as explained in Ref~\cite{ArielPRA20}. Thus, in order to extend the applicability of the proposed approach we also considers a serial concatenation of AD and PD.

The Kraus operators of this combined channel are obtained by multiplication of the Kraus operators of AD and PD channels \cite{Wilde}
\begin{equation}
K^{(ap)}_k (t)=K^{(a)}_i (t) K^{(p)}_i (t),
\end{equation}
where $i,j=0,1$, $k=0,1,2,3$, and $K^{(a)}_i$ and $K^{(p)}_j$ are given in Eqs.~\eqref{eq:kraus_AD} and \eqref{eq:kraus_PD}, respectively. We note that the resultant density matrix of the evolved qubit 
is independent of the order of AD and PD channels, i.e., $\rho^{(ap)}(t)=\rho^{(pa)}(t)$.

This combined channel can be simulated with a quantum circuit that contains two ancillae, one for PD and one for AD, with the corresponding gates for each channel acting on the qubit and each ancilla. 

\subsection{Non-markovianity measure}\label{sec:NMmeasure}

Several measures of non-Markovianity have been introduced \cite{Chruscinski,Breuer,Luo,PollockPRL18, Rivas}. In this work, we consider the measure based on entanglement dynamics of a bipartite quantum state. This bipartite system is composed of the system and an auxiliary ancilla that is isolated from the environment \cite{Rivas}. It is worth noticing that this ancilla only serves the theoretical purpose of quantifying non-Markovianity and it is not implemented in the quantum circuits. On the contrary, the ancilla (qubit a) is present in the quantum circuit since it is used to simulate the effect of the environment.

A monotonic decrease in the entanglement of the bipartite system implies that the dynamics is Markovian. An increase in the entanglement during the evolution is a result of memory effects and thus non-Markovianity. Using this criteria, the degree of non-Markovianity can be calculated as 
\begin{equation}
    \mathcal{N}=\max_{\rho(0)}\int_{dE(t)/dt>0}{\frac{dE(t)}{dt}dt},
\end{equation}
where the maximization is done over all initial states $\rho(0)$ and $E(t)$ is the measure of entanglement where we use concurrence \cite{Hill}. It has been found that the maximization is achieved for Bell states \cite{Cervati}. Therefore, we consider a bipartite system in a Bell state.

In the following sections, we first briefly review VQCs. We then use VQCs to find an optimal sequence to estimate the degree of non-Markovianity of a qubit under PD, AD and the combination of both dissipative channels.


\section{Variational quantum circuit}\label{QVC_sec}
VQCs, also known as parametrized quantum circuits (sometimes also referred to as quantum neural networks), are a type of quantum circuit with some free parameters (normally for single qubit operations) \cite{Benedetti}. We label the free quantum parameters of the circuit by $\varphi$. In addition, we consider classical parameters, labelled by $w$, in post-processing the measurement outcomes of the quantum circuit.

The optimization of both quantum and classical parameters $\varphi$ and $w$ can be done in a supervised manner using a set of input data, $x$, and their corresponding labels, $f_t(x)$. We aim to minimize a generic cost function defined as the mean square error (MSE), i.e.,

\begin{equation}\label{eq:costfun}
    C(x,\varphi,w)=\frac{1}{n}\sum_{i=1}^{n}\left({\check{f}_{\varphi,w} (x_i)-f_t (x_i)}\right)^2,
\end{equation}
where $n$ is the number of input data. In this approach, $\check{f}_{\varphi, w}(x)$ is the estimate of $f_t(x)$ obtained from the output of the VQC. The output $\check{f}_{\varphi, w}(x)$ is calculated as a linear combination of the measurement outcomes $\langle M_i \rangle_{x,\varphi}$: 
\begin{equation}\label{eq:vqcout}
  \check{f}_{\varphi, w}(x)=w_0 +\sum_{i=1}^{k}{ w_i \langle M_i \rangle_{x,\varphi}}.
\end{equation}
Here, $k$ is the number of measurement outcomes performed on the circuit. The whole optimization can be done through gradient-based techniques, such as stochastic gradient descent \cite{Robbins}, gradient-free techniques such as Nelder-Mead method \cite{Nelder}, or particle swarm optimization \cite{Zhu}. The gradient can also be calculated using the quantum circuit \cite{SchuldPRA19}.

\section{Estimation of the degree of non-Markovianity}\label{regg_sec}
\begin{figure*}[t!]
\centering
    \includegraphics[width=1\textwidth]{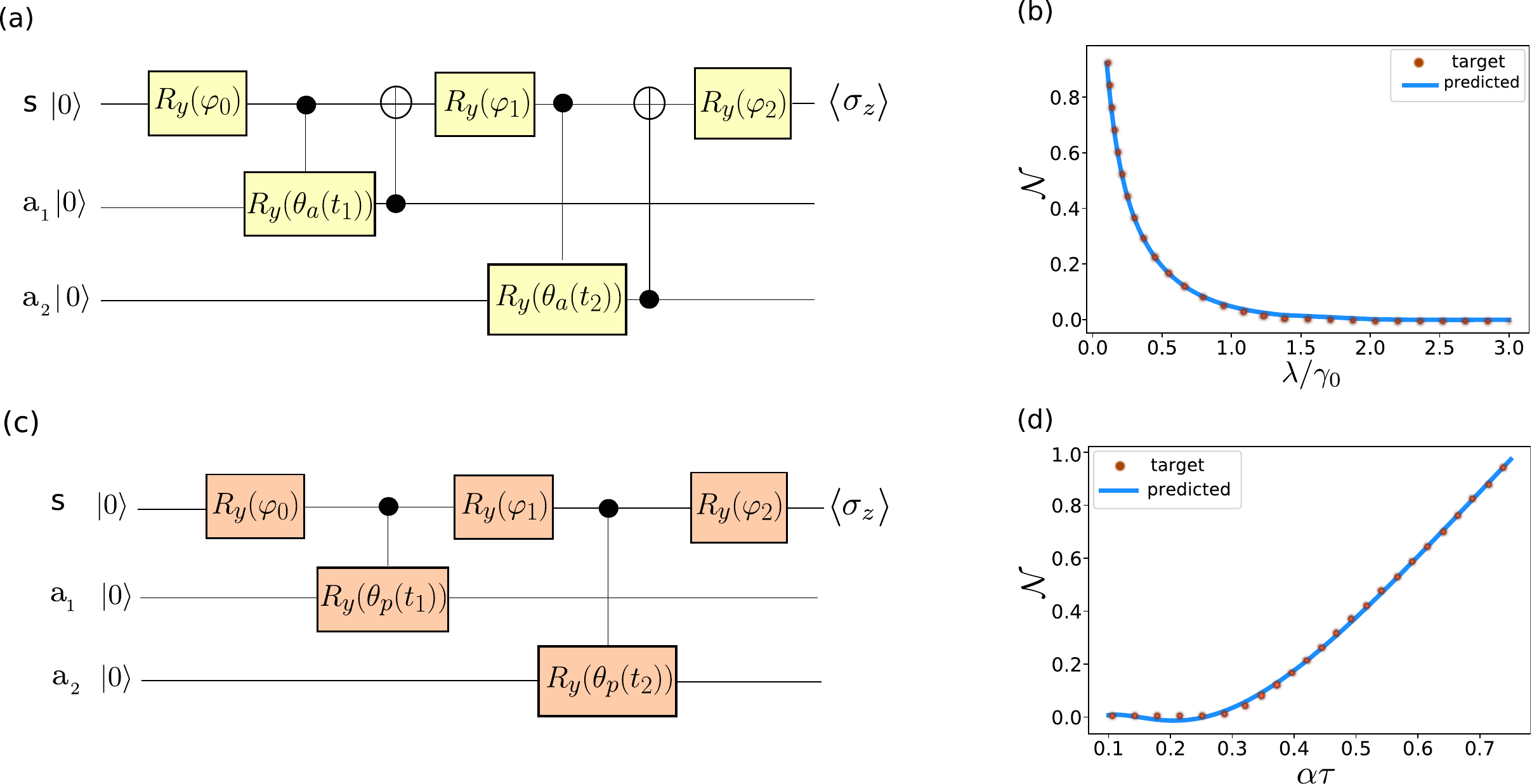}
   \caption{VQCs for estimating the degree of non-Markovianity $\mathcal{N}$ for AD channel ({\it a}) and PD channel (c). In each circuit the system qubit ({\it s}) is first initialized by the rotation $R_y (\varphi_0)$. The qubit {\it s} then interacts with the environment in a sequence of two interactions with time $t_i$. After each interaction a rotation $R_y (\varphi_i)$ acts on the qubit {\it s}. The two ancilla qubits (a$_1$ and a$_2$) simulate the effect of the environment for each interaction, one ancilla for each interaction. In the output, the qubit {\it s} is measured in $\sigma_z$ basis. (b) and (d) show the predicted (solid blue line) and target values (dotted red line) of $\mathcal{N}$ versus $\lambda/\gamma_0$ for AD and versus $\alpha \tau$ for PD, respectively.}
   \label{fig:ADPDreg}
\end{figure*}
We use VQCs to estimate the degree of non-Markovianity $\mathcal{N}$ of the dynamics of a qubit, under AD and PD channels. We first consider each of these channels independently. We found that a precise estimate of $\mathcal{N}$ can be obtained using a sequence of qubit-environment interactions. Our proposed scheme is the following: the qubit, initially in the state $|0\rangle$, is rotated along the $y$ axis with angle $\varphi_0$. Next, it interacts with the environment in a sequence of two interactions, each with time $t_i$. Each interaction is followed by a rotation along the $y$ axis with angle $\varphi_i$ ($i=1,2$). Finally, the qubit is measured in the Pauli $z$ basis. The estimate of $\mathcal{N}$ is obtained as
\begin{equation}\label{eq:Ncheck}
    \check{\mathcal{N}}=w_0+w_1\langle\sigma_z\rangle.
\end{equation}

We consider $\varphi_i$ ($i=0,1,2$), $t_j$ ($j=1,2$), $w_0$, and $w_1$ as the training parameters. The VQCs for AD and PD channels, are shown in Fig.~\ref{fig:ADPDreg}(a) and (c), respectively. The optimal values of these parameters are found by minimizing the cost function, Eq.~\eqref{eq:costfun}, for a range of values of the parameters $\lambda/\gamma_0$ (for AD) and $\alpha\tau$ (for PD). Each VQC consists of a qubit (s) and two ancillas (a$_1$ and a$_2$). Each ancilla is used to simulate the dynamics of the qubit in each of its interactions with the environment. Note that, in the experiment, only one qubit (that interacts with the environment) is required to perform this estimation task, and no operation is required to be performed on the environment. 

We remark that the selection of this ansatz for the structure of the circuit is influenced by the evolution of the Bloch vector and our intuition about the separation of two classes, \textit{i.e.} Markovianity and non-Markovianity. We refer to this choice as a problem-based ansatz, although the approach could be extended to other tasks. 

We performed our simulations using the quantum simulator of Pennylane \cite{Pennylane}. We used a combination of RMSprop \cite{Hinton} and Adagrad optimizers \cite{Duchi} (see Appendix \ref{appendix:optimizer} for some details about the optimizers). 
In order to find the optimal values of the parameters we generate 1000 data points (feature vectors) for each channel. The data is uniformly distributed for the parameter $\lambda/\gamma_0$ in the range $[0.1, 3]$ (for AD) and for the parameter $\alpha\tau$ in the range $[0.1, 0.75]$ (for PD). The feature vectors contain the values of the parameters $\lambda/\gamma_0$ (AD) and $\alpha \tau$ (PD) and the labels are the corresponding values of $\mathcal{N}$ for each channel. 

We use 50$\%$ of the whole data points (which are chosen randomly) to train the VQC for each channel. For AD channel, we obtained $7.2\times 10^{-6}$ and for the PD channel we obtained $4.3\times 10^{-5}$ for the MSE, testing on the whole data points. Figures ~\ref{fig:ADPDreg}(b) and (d) show the target (dotted red line) and the predicted values (solid blue line) of $\mathcal{N}$ versus $\lambda/\gamma_0$ (for AD) and versus $\alpha \tau$ (for PD), respectively. For the AD channel, the predicted values are very close to the target values for the full range of $\lambda/\gamma_0$. For the PD channel, there is a small difference between the target and predicted values of $\mathcal{N}$ where $\mathcal{N}=0$.

Using only one interaction of the qubit with the environment, for the AD channel, we obtained $1\times 10^{-4}$ for the MSE, while for the PD channel, we were not able to obtain a good estimate of $\mathcal{N}$. A single interaction of the qubit with the environment is a nonlinear mapping from the characteristic parameter of the environment ($\lambda/\gamma_0$ for AD and $\alpha \tau$ for PD) to the state of the qubit, which is measured by $\langle\sigma_z\rangle$. For the case of AD, one interaction (with a properly chosen interaction time) is sufficient to obtain a reasonable estimate of $\mathcal{N}$, while for the case of PD, only one interaction is not sufficient. After the first interaction with the environment, each value of the characteristic parameter is mapped to a different Bloch vector, from which the nonlinear mapping of the second interaction gives $\mathcal{N}$ with high precision. It is worth mentioning that, for the case of PD, we did not obtain a significant decrease in MSE by increasing the number of interactions of the qubit with the environment (up to 5 interactions).
\begin{figure*}[t!]
\centering
    \includegraphics[width=1\textwidth]{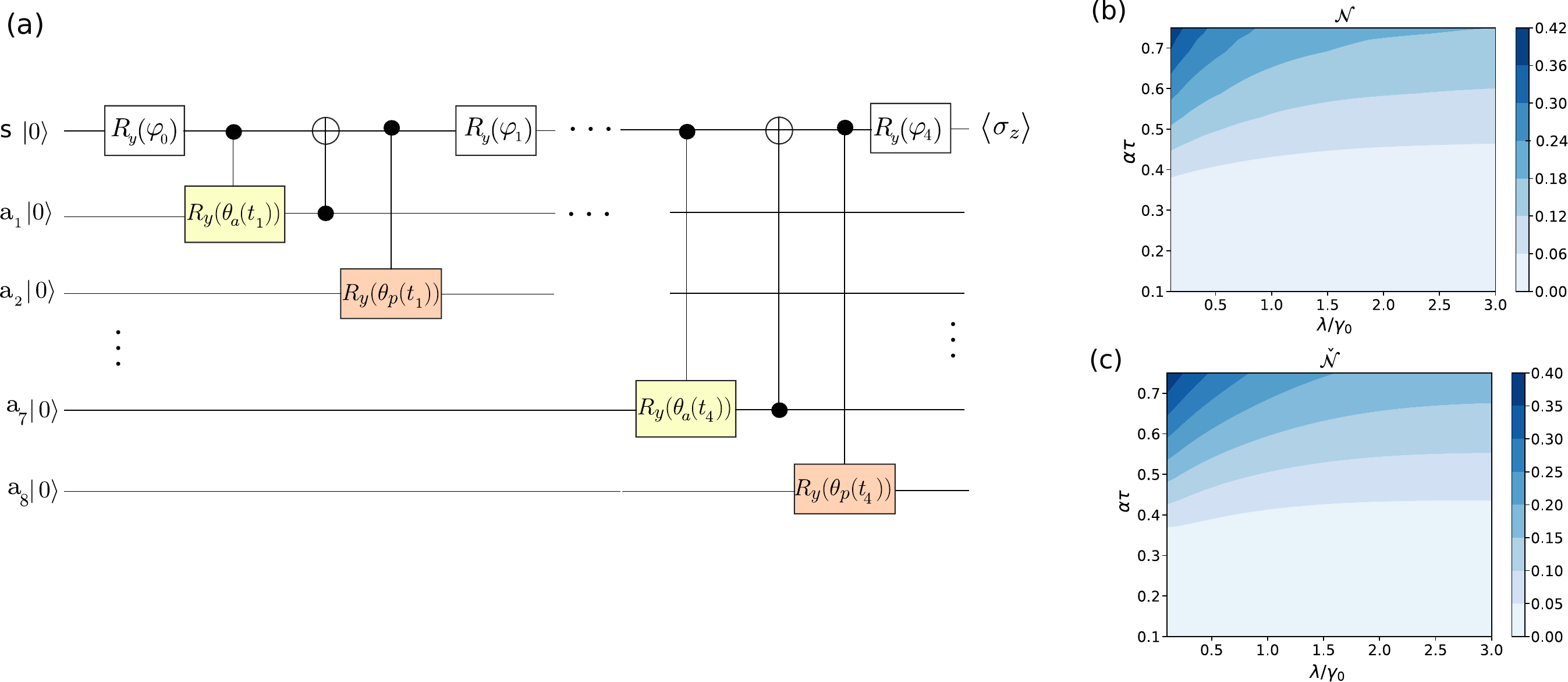}
   \caption{(a) VQC for estimating the degree of non-Markovianity $\mathcal{N} $ for the case of combined amplitude and phase damping. The system qubit ({\it s}) is first initialized by the rotation $R_y (\varphi_0)$. The qubit {\it s} then interacts with the environment in a sequence of four interactions with time $t_i$. After each interaction a rotation $R_y (\varphi_i)$ acts on the qubit {\it s}. For each interaction, two ancilla qubits are used to simulate the effect of the environment, one for AD and one for PD. In the output, the qubit {\it s} is measured in $\sigma_z$ basis. (b) and (c) show the contour plots of the degree of non-Markovianity, $\mathcal{N}$, and $\check{\mathcal{N}}$ (estimate of $\mathcal{N}$) versus $\lambda/\gamma_0$ and $\alpha \tau$, respectively.}
   \label{fig:ADyPDreg}
\end{figure*}

We also considered serial concatenation of AD and PD channels. In this case, we considered $2025$ data points in the range [0.1,3] for $\lambda/\gamma_0$ and $[0.1,0.75]$ for $\alpha \tau$ (45 uniformly distributed data points for each range), and defined the estimate of $\mathcal{N}$ as
\begin{equation}
\check{\mathcal{N}}=\text{max}\left(w_0+w_1\langle\sigma_z\rangle,0\right)
\end{equation}
to avoid negative values. Training of the circuit is done using 70$\%$ of the data points. We obtained a precise estimate of $\mathcal{N}$ ($7.8\times 10^{-6}$ testing on the whole dataset) using a sequence of four interactions of the qubit with the environment. As the dynamics (in this case) depends on two parameters, $\alpha \tau$ and $\lambda/\gamma_0$, a larger number of interactions of the qubit with the environment is required to achieve a precise estimate. After each combined AD and PD interaction a $R_y(\varphi)$ gate is applied on the qubit (Fig.~\ref{fig:ADyPDreg}a). The optimization is done over the interaction times $t_i$, $i=1,...,4$, the rotation angles $\varphi_j$, $j=0,...,4$, and the parameters $w_0$ and $w_1$. Note that $R_y (\varphi_0)$ is the initialization gate and we have taken the interaction time of AD and PD channels to be equal.

\section{Conclusions}\label{conc_sec}
In summary, we proposed an experimentally feasible scheme to estimate the degree of non-Markovianity based on entanglement dynamics. We implemented this approach to the paradigmatic models of phase, amplitude damping and combination of both. We assumed that the type of decoherence channel and the range of characteristic parameters of the channel that determine the non-Markovianity of the process are known. In contrast to previous works, the methodology proposed in this work uses variational quantum circuits (through supervised learning) to take advantage of the quantum nature of the problem. Based on our physical knowledge we provided a problem-based ansatz that involves a sequence of qubit-environment interactions yielding accurate estimations of the degree of  non-Markovianity from measurement outcomes with minimal classical post-processing. Finally, the ansatz proposed here for the variational quantum circuit may be used for other estimation and classification tasks. 

\section{Acknowledgement}
The authors thank Mauro Cirio for helpful comments on the manuscript. H.T.D. acknowledges support from Universidad Mayor through a postdoctoral fellowship. D.T acknowledges financial support from Universidad Mayor through the Doctoral fellowship. F.F.F. acknowledges support from Funda\c{c}\~ao de Amparo \'a Pesquisa do Estado de S\~ao Paulo (FAPESP), Project No. 2019/05445-7. A.~N.~acknowledges financial support from Fondecyt Iniciaci\'{o}n No 11220266. R.C. acknowledges financial support from FONDECYT Iniciaci\'on No. 11180143.


\appendix
\section{Optimization algorithms}\label{appendix:optimizer}
In this appendix, following Ref.~\cite{Ruder} we give some details about the optimizers that we used in our simulations.
\subsection{Adagrad}
Adaptive gradient (Adagrad) is a variation of the gradient descent (GD) optimizer. In the GD algorithm, the parameters, labelled by $\varphi$ here, are updated in the opposite direction of the gradient of the cost function $C(\varphi)$. In other words, for every parameter $\varphi_i$ at each time step $t$ the update rule can be written as
\begin{equation}
    \varphi_{t+1,i}=\varphi_{t,i}-\eta \nabla_{\varphi_i}C(\varphi),
\end{equation}
where $\eta$ is the learning rate, which is assumed to be constant and independent of $\varphi_i$ thorough the learning process. 

In Adagrad a different learning rate is used for every parameter $\varphi_i$ at every time step. In the update rule for Adagrad, the learning rate at each time step $t$ for every parameter $\varphi_i$ is based on the past gradients that have been calculated for $\varphi_i$ \cite{Ruder}
\begin{equation}
    \varphi_{t+1,i}=\varphi_{t,i}-\frac{\eta}{\sqrt{g_{t+1,i}+\varepsilon}}\nabla_{\varphi_i}C(\varphi).
\end{equation}
Here, $g_{t+1,i}$ is the sum of the squares of the gradients with respect to $\varphi_i$ up to time step $t+1$,
\begin{equation}
g_{t+1,i}=g_{t,i}+\left(\nabla_{\varphi_i}C(\varphi)\right)^2,
\end{equation}
where $g_0=0$, and $\varepsilon$ (usually chosen on the order of $10^{-8}$) is for avoiding division by zero.

The weakness of adagrad is the accumulation of the squared gradients in the denominator. The accumulated sum keeps growing during the training process as the added terms are all positive. As a result, the learning rate could approach zero and therefore the algorithm stops learning. To avoid this issue in our simulations, when the rate of learning became very small we reinitialized the optimization process with the newly found hyperparameters as the initial values.

\subsection{RMSprop}
Root mean square propagation (RMSprop) is a variation of Adagrad algorithm which uses a decaying average of squared gradients in the adaptation of the step size for each parameter \cite{Ruder}. The use of a decaying average allows the algorithm to forget early gradients and only focus on the most recent gradients during the optimization process. As a result, RMSprop overcomes the AdaGrad's diminishing learning rates. In RMSprop, the parameter update rule is
\begin{equation}
    \varphi_{t+1,i}=\varphi_{t,i}-\frac{\eta}{\sqrt{g_{t+1,i}+\varepsilon}}\nabla_{\varphi_i}C(\varphi).
\end{equation}
In this case, we have
\begin{equation}
g_{t+1,i}=\gamma g_{t,i}+(1-\gamma)\left(\nabla_{\varphi_i}C(\varphi)\right)^2,
\end{equation}
where $g_{0}=0$, and for $\gamma$ it is suggested to use $\gamma=0.9$ \cite{Hinton}.

In our simulations, using RMSprop algorithm a local minima of the cost function reached quickly (in less number of steps than Adagrad). However, the algorithm then started to diverge, resulting in large values for the cost function. Therefore, once RMSprop reached the local minima, we followed the optimization process with Adagrad to reach the desired accuracy.

\end{document}